\documentclass[10pt]{article}

\usepackage{graphicx,amsmath,amssymb,url}

\newlength{\dinwidth}
\newlength{\dinmargin}
\def\lapproxeq{\lower .7ex\hbox{$\;\stackrel{\textstyle
<}{\sim}\;$}}
\def\gapproxeq{\lower .7ex\hbox{$\;\stackrel{\textstyle
>}{\sim}\;$}}
\def\gtrsim{\lower .7ex\hbox{$\;\stackrel{\textstyle
>}{\sim}\;$}}
\def\lesim{\lower .7ex\hbox{$\;\stackrel{\textstyle
<}{\sim}\;$}}

\def\be{\begin{equation}}
\def\ee{\end{equation}}
\def\bea{\begin{eqnarray}}
\def\eea{\end{eqnarray}}

\def\J{J/\psi}

\def\MS{\overline{\rm MS}}

\begin{document}

\titlepage

\begin{flushright}
MPP-2016-301\\
IPPP/16/84\\
LTH 1100\\

\today\\

\end{flushright}

\vspace*{0cm}

\begin{center}

{\Large \bf Exclusive $J/\psi$ process tamed to probe the low $x$ gluon\footnote{Talk presented by A.D.Martin at Diffraction 2016, Acireale, Sicily, Sept. 2-8, 2016, to be published in AIP conference Proceedings.}}

\vspace*{1cm} {\sc S.P. Jones}$^{a}$, {\sc A.D. Martin}$^b$,  {\sc
  M.G. Ryskin}$^{b,c}$ and {\sc T. Teubner}$^{d}$ \\

\vspace*{0.0cm}
$^a$  {\em Max-Planck-Institute for Physics, Fohringer Ring 6, 80805 Munchen, Germany}\\
$^b$ {\em Institute for Particle Physics
  Phenomenology, Durham University, Durham DH1 3LE, U.K.}\\
$^c$ {\em Petersburg Nuclear Physics Institute, NRC Kurchatov Institute, Gatchina,
St.~Petersburg, 188300, Russia} \\
$^d$ {\em Department of Mathematical Sciences,
University of Liverpool, Liverpool L69 3BX, U.K.}\\
 \end{center}

\vspace*{0cm}

\begin{abstract}

We address the question as to whether data for $J/\psi$ mesons produced exclusively in the forward direction at the LHC can be used in global parton analyses (based on collinear factorization) to pin down the low $x$ gluon PDF.   We show that it may be possible to overcome the problems that (i) the process is described by a skewed or Generalized Parton Distribution (GPD), (ii) it is very sensitive to the choice of factorization scale and (iii) there is bad LO, NLO,... perturbative stability to the predictions.  However, we start by briefly explaining how the alternative $k_T$ factorization approach has been used to describe the process.

\end{abstract}

\vspace*{0.0cm}

\section{Introduction}

\begin{figure} [h]
\begin{center}
 \includegraphics[clip=true,trim=1.0cm 5.0cm 0.0cm 6.0cm,width=12.0cm]{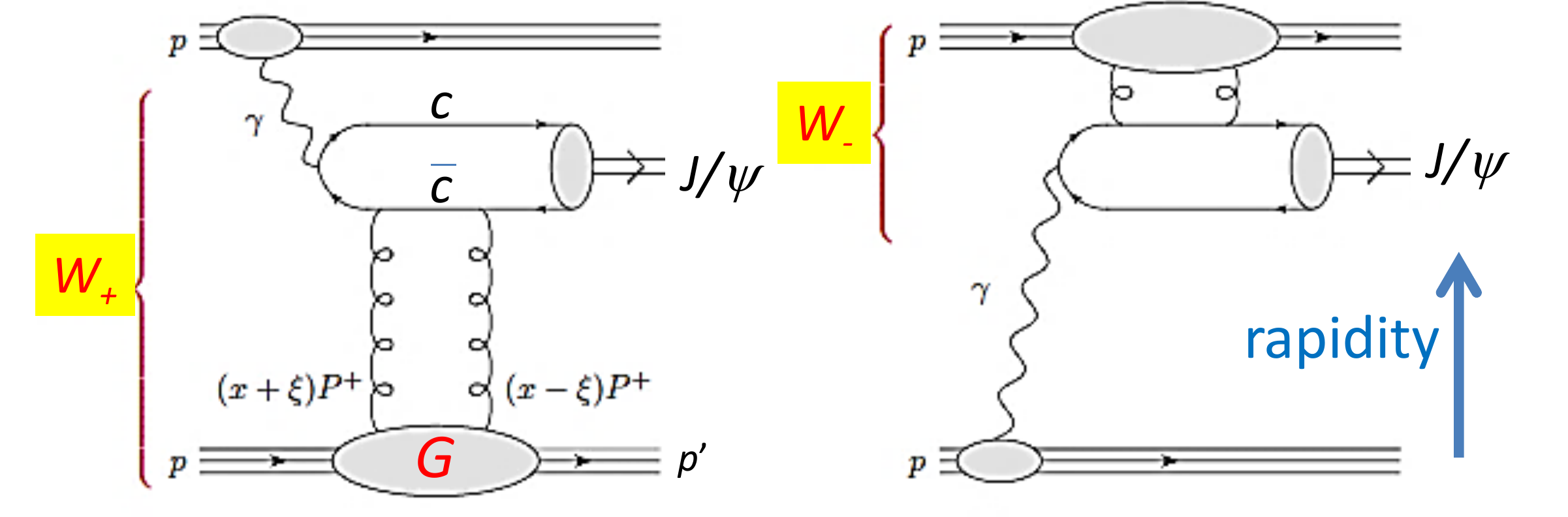} 
 \caption{\sf $d\sigma(pp\to p+\J+p)/dy$ driven by the subprocess $\gamma p\to \J+p$ at two different energies, $W_\pm$. }
\label{fig:subproc}
\end{center}
\end{figure}

\begin{figure} [h]
\begin{center}
 \includegraphics[clip=true,trim=.0cm 7.0cm 0.0cm 3.5cm,width=17.0cm]{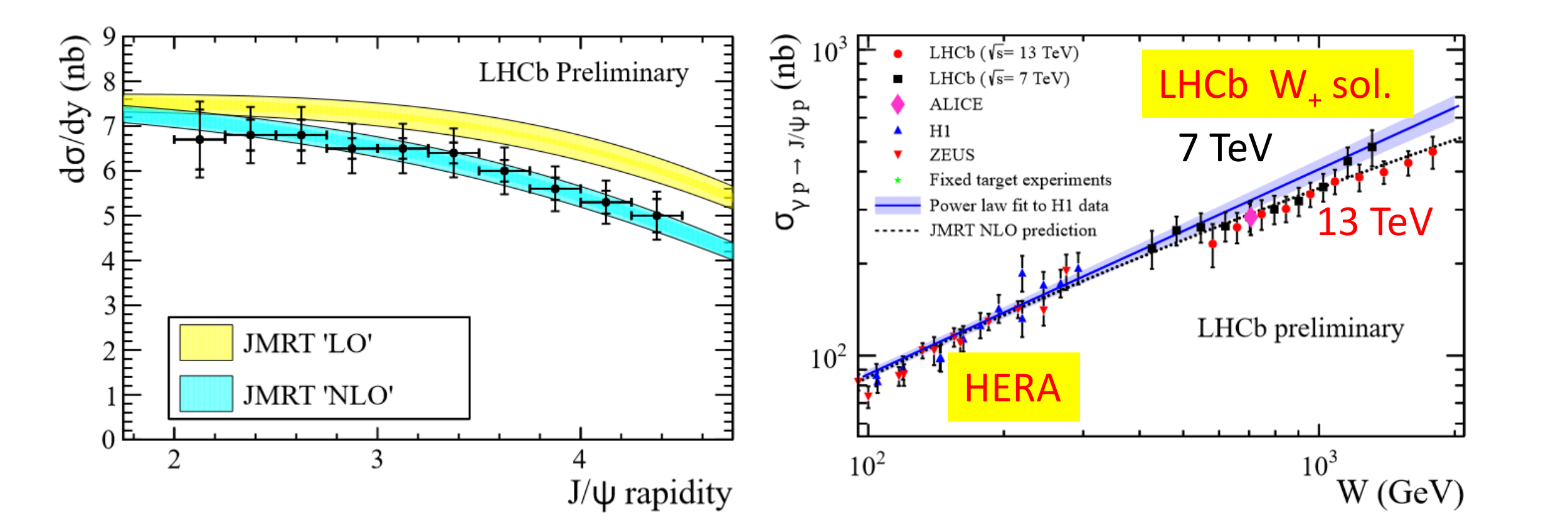} 
 \caption{\sf Left plot: LHCb data for $d\sigma(pp\to p+\J+p)/dy$ \cite{LHCb16}. Right plot: the resulting beaviour of $\sigma(\gamma p\to \J+p)$ as a function of the $\gamma p$ energy, $W$. The plots are taken from Ref.~\cite{LHCb16}.}
\label{fig:LHCb16}
\end{center}
\end{figure}

As we shall see, LHCb data for the exclusive process $pp \to p~+~J/\psi~+p$ in the rapidity interval $2<y(J/\psi)<4.5$ should, in principle, be able to probe the gluon PDF down to about $x= (M_\psi /\sqrt{s})~
e^{-y}\sim 10^{-5}$.
The process is driven by the quasielastic subprocess $\gamma^*p\to J/\psi~+~p$, see Fig.~\ref{fig:subproc}.
In fact LHCb data for this process at 13 TeV have just become available \cite{LHCb16}, see the first plot in Fig.~\ref{fig:LHCb16}. 

Following Fig.~\ref{fig:subproc}, the LHCb collaboration have extracted the cross sections for $\gamma p\to J/\psi~+~p$ from the $pp$ data  using
\be
\frac{d\sigma(pp)}{dy}~=~S^2(W_+)\left(k_+\frac{dn}{dk_+} \right)\sigma_+(\gamma p)~+~S^2(W_-)\left(k_-\frac{dn}{dk_-} \right)\sigma_-(\gamma p)
\ee
where the survival probabilities of the large rapidity gaps, $S^2$, and the photon fluxes, $k dn/dk$, are known; $k$ is the energy of the photon. There are two contributions according as to whether the photon is emitted from one or the other proton, with corresponding different $\gamma p$ energies squared $W^2_\pm =M_\psi \sqrt{s} ~e^{\pm |y|}$. The interference term is negligible.

Long ago, Ryskin \cite{Ryskin} gave the LO expression for the exclusive cross section in terms of the gluon PDF 
\begin{equation}
\frac{{\rm d}\sigma}{{\rm d}t}\left( \gamma p \to \J~ p \right)
     {\Big |}_{t=0} = \frac{\Gamma_{ee}M^3_{\psi}\pi^3}{48\alpha}\,
     \left( 1+\frac{Q^2}{M^2}\right)
\left[\frac{{\alpha_s(\bar Q^2)}}{\bar Q^4}xg(x,\mu_F^2)\right]^2\,,
\label{eq:lo}
\end{equation}
with $\mu_F\sim \bar{Q}$, and where $M_{\psi}$ and $\Gamma_{ee}$ are the mass and electronic
width of the $\J$. The kinematic variables are 
\begin{equation}
{\bar Q^2}~=~(Q^2+M^2_{\psi})/4\,, ~~~~~~~~~x~=~(Q^2+M^2_{\psi})/(Q^2+W^2)\,,
\end{equation}
and $W$ is the $\gamma p$ centre-of-mass energy. We assume
the $t$ dependence to be exponential, i.e. $\sigma =
\exp(-Bt)$, where the energy-dependent $t$ slope parameter, $B$, has the form 
\begin{equation}
B(W) = \left(4.9 + 4 \alpha' \ln(W/W_0)\right) {\rm\
  GeV}^{-2}\,,
\label{eq:b-slope}
\end{equation}
where the pomeron slope $\alpha'=0.06$ GeV$^{-2}$ and $W_0=90$~GeV.  From Fig.~\ref{fig:subproc} we see that the quasielastic process actually depends on the gluon Generalized Parton Distribution (GPD$(\xi,x)$),  where $\xi=(p^+-p^{\prime +})/(p^++p^{\prime+})$ is the skewedness parameter of Fig.~\ref{fig:subproc}. However this is not a problem, since for $\xi < |x|\ll 1$
\be
{\rm GPD}(\xi,x)~=~{\rm PDF}(x') \otimes {\rm Shuvaev}(\xi,x,x'),
\ee
to $\mathcal{O}(\xi)$, where the conventional PDF is convoluted with the Shuvaev tranform  \cite{Shuvaev}.
We will allow for skewing and the real part of the amplitude exactly as in \cite{JMRT1307}.  

To constrain the \textbf{collinear factorization} scale, $\mu_F$, at which the gluon is measured, we need the NLO correction. However, we encounter very bad convergence of the LO, NLO,... perturbation series at low $\xi$ and low scales.  The reason can be seen by estimating the average number $\langle n \rangle$ of additional gluons emitted in the low $\xi,~\mu_F$ domain
\be
\langle n \rangle~\simeq~(3\alpha_S/\pi)~{\rm ln}(1/\xi) ~\Delta{\rm ln}(\mu^2_F)~\sim ~5,
\ee
whereas including the NLO correction accounts for only one additional gluon!

\section{$k_T$ factorization approach}

\begin{figure} [h]
 \begin{center}
 \includegraphics[clip=true,trim=0.0cm 7.cm -.9cm 1.0cm,height=6.0cm]{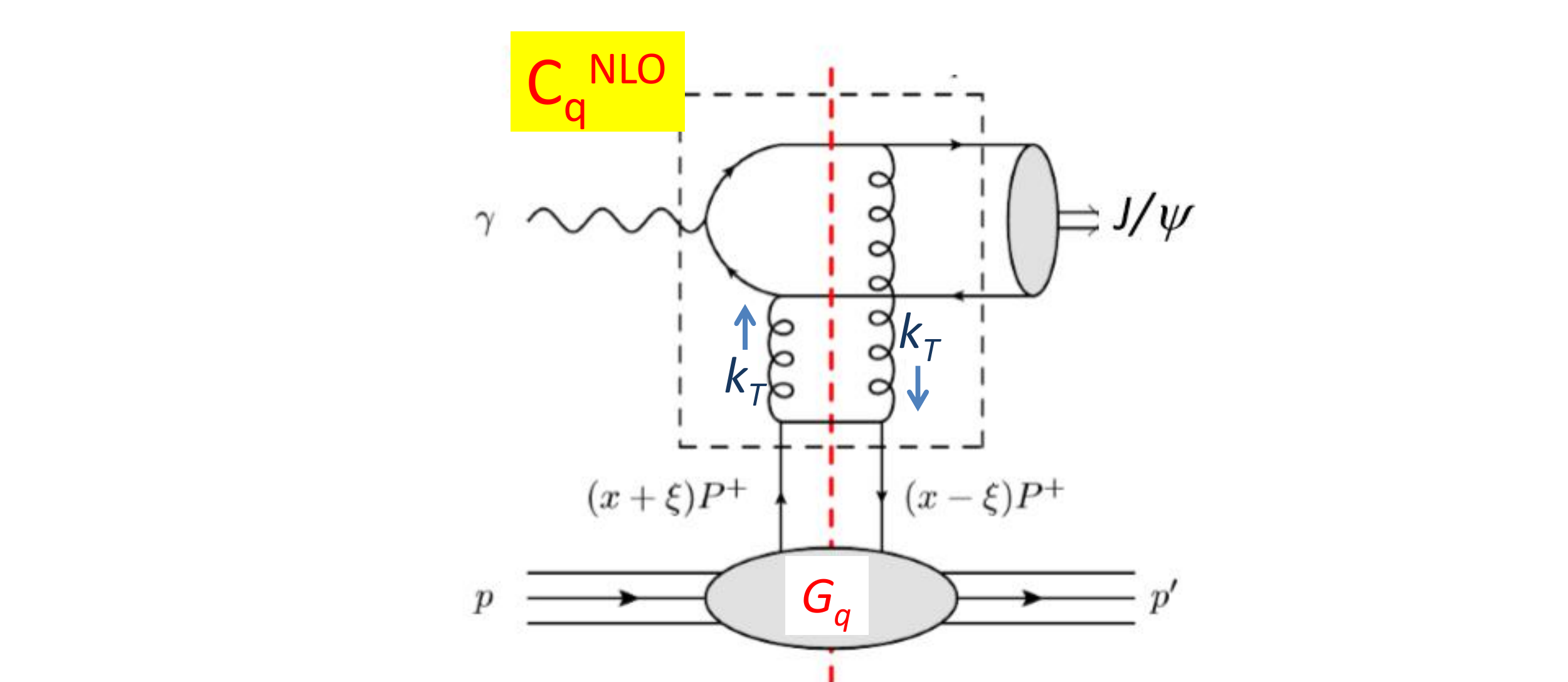} 
 \caption{\sf The $k_T$ integration performed on the last step of the evolution to obtain the `NLO' quark coefficient function $C^{\rm NLO}_q$. The lower quark line is replaced by a gluon to obtain the `NLO' gluon coefficient function $C^{\rm NLO}_g$.}
\label{fig:cfnq}
\end{center}
\end{figure}

Given the above remarks, why is the JMRT `NLO' prediction so reasonable in Fig.~\ref{fig:LHCb16}?  The reason is that we use the \textbf{$k_T$ factorization} procedure to obtain the approximate NLO correction to the coefficient functions by performing the explicit $k_T$ integration in the last step of the evolution \cite{laststep}, and use an input PDF with resummed $(\alpha_S{\rm ln}(1/\xi){\rm ln}(\mu_F^2))^n$ terms arising from ladder diagrams, see Fig~\ref{fig:cfnq}. This is not the complete NLO contribution, but it includes the most important diagrams at low $x$ and low $\mu^2_F$.  To do this we need the gluon PDF unintegrated over $k_T$,
\be
f(x,k_T^2)~=~\partial[xg(x,k^2_T)T(k^2_T,\mu^2)]/\partial{\rm ln}k^2_T,
\ee
where $\mu^2={\rm max}(k^2_T,\bar{Q}^2)$, and where  the Sudakov factor $T$ is required to ensure no additional gluons are emitted with transverse momenta greater than $k_T$.
That is, we replace the [....] in (\ref{eq:lo})  by 
\be
\left[\frac{{\alpha_s(\bar Q^2)}}{\bar Q^4}xg(x,\bar Q^2)\right]
\:\longrightarrow\: 
\int_{Q_0^2}^{(W^2-M_{\psi}^2)/4} 
\frac{{\rm d}k_T^2\,\alpha_s(\mu^2)}{\bar Q^2 (\bar Q^2 + k_T^2)} \, 
\frac{\partial \left[ xg(x,k_T^2) \sqrt{T(k^2_T,\mu^2)}
  \right]}{\partial k_T^2}  ~~~+ Q_0 ~{\rm contribution},
\label{eq:nlointegral}
\ee
where the convergence of the integral over $k_T$ is ensured (even for an infinite upper limit)  by the factor $1/(\bar{Q}^2+k_T^2)$.
By parameterizing the gluon $xg(x,\mu^2)$ in double logarithm form we sum the leading $(\alpha_s{\rm ln}(1/\xi){\rm ln}(\mu_F^2))^n$ contributions.  The parameters were obtained \cite{JMRT1307} by fitting the 7 TeV LHCb data \cite{LHCb13}, and were used to make the predictions shown in Fig.~\ref{fig:LHCb16}.

\section{Taming~ NLO~ in ~the ~collinear ~scheme}

\begin{figure} [h]
\begin{center}
 \includegraphics[clip=true,trim=0.5cm 5.5cm 0.0cm 5.cm,height=6.cm]{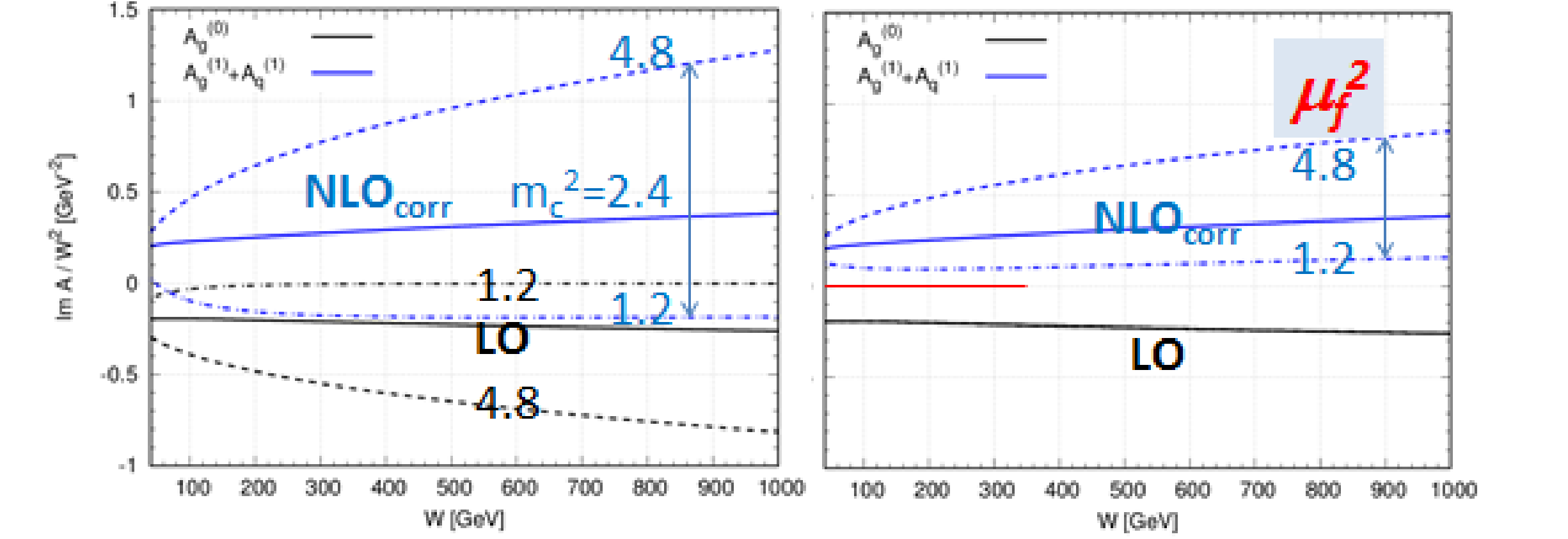} 
 \caption{\sf Left plot: predictions of Im$A/W^2$ for $\gamma p\to \J+p$ for different values of the collinear factorization scale, namely $\mu_F^2=4.8,~2.4,~1.2$ GeV$^2$. Right plot: the predictions after the transfer of the $(\alpha_s{\rm ln}(1/\xi){\rm ln}(\mu_F^2))^n$ terms from the NLO coefficient function to the LO contribution by the particular choice $\mu_F^2=m_c^2=2.4$ GeV$^2$; we have less dependence on the new factorization scale $\mu_f$. We use CTEQ6.6 PDFs \cite{CTEQ} to be consistent with earlier work and to ensure a positive gluon PDF at low $x$ and low $Q^2$.}
 \label{fig:scale}
 \end{center}
 \end{figure}

Unfortunately the `NLO' gluon PDF obtained in the $k_T$ factorization scheme, cannot be directly related to the $\MS$ PDFs of the global parton analyses. There we work in the collinear factorization scheme. Although the NLO contribution is explicitly known in this scheme, there are problems \cite{Ivanov,JMRT1507}. As mentioned before, there is very bad perturbative convergence in the prediction for $\gamma p \to \J+p$. Indeed, the NLO correction is comparable or larger than the LO result and is opposite in sign. Moreover there is strong dependence on the choice of the factorization scale $\mu_F$. This is clearly visible in the left plot of Fig. \ref{fig:scale}, which shows the predictions at LO, and the correction due to NLO, for factorization scales $\mu_F=4.8,~2.4,~1.2$ GeV$^2$.  However we can improve the situation by resumming the $(\alpha_s{\rm ln}(1/\xi){\rm ln}(\mu_F^2))^n$ terms and moving them into the LO contribution by a particular choice of factorization scale; namely $\mu_F=m_c$. The details are given in Ref.~\cite{JMRT1507}. The result is that the $\gamma p \to \J+p$ amplitude is of the form
\be
A(\mu_f)~=~C^{\rm LO} \otimes {\rm GPD}(\mu_F)~+~C^{\rm NLO}_{\rm rem}(\mu_F)\otimes{\rm GPD}(\mu_f).
\ee
With this choice of $\mu_F$ there is a smaller remaining contribution in the NLO coefficient function, and so the residual dependence on the scale $\mu_f$ is reduced, as seen in the right plot of Fig. \ref{fig:scale}. Nevertheless we still have very bad perturbative convergence. The NLO correction is still comparable to the LO result, and opposite in sign!

\begin{figure} [h]
 \begin{center}
 \includegraphics[clip=true,trim=0.0cm .cm .cm 1.0cm,height=8.5cm]{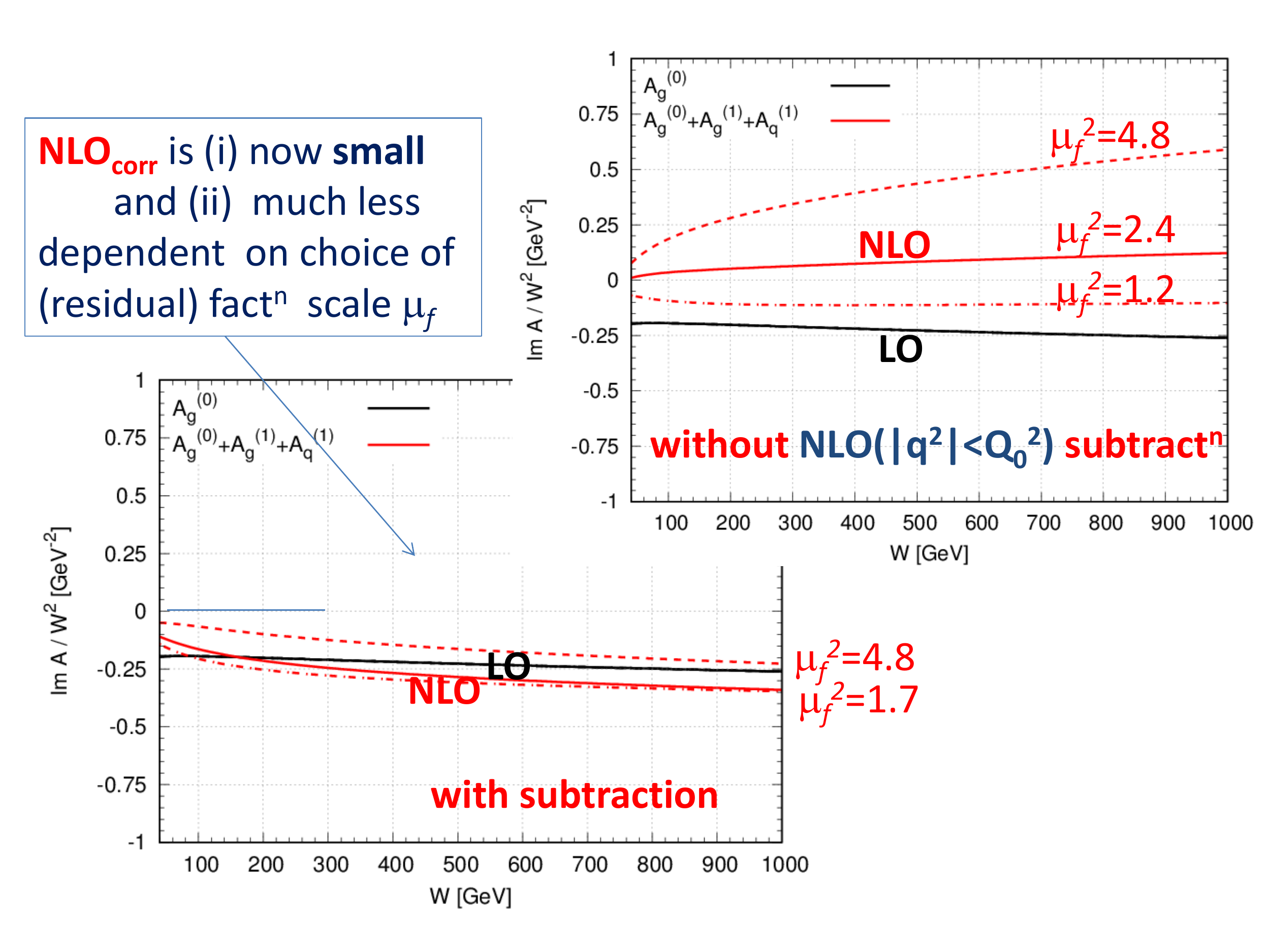} 
 \caption{\sf The upper and lower plots are respectively Im$A/W^2$ for $\gamma p\to \J+p$ before and after the $Q_0$ subtraction has been performed. The lower plot shows much less sensitivity to the new factorization scale $\mu_f$, and that reasonable perturbative stability has been achieved. }
\label{fig:Capture}
\end{center}
\end{figure}

Can anything more be done? Yes. We must investigate the effect of an important $Q_0$ cut. Recall DGLAP evolution starts at some input scale $Q_0$. At leading order everything below $Q_0$ is included in the input PDFs at $Q_0$.  However, at NLO, the contribution to the coefficient functions from the region $|q^2|<Q^2_0$ result in double counting. Here $q$ is the four momentum of the $t$-channel gluons in  the collinear version of the quark coefficient function of Fig.~\ref{fig:cfnq}. To be consistent we need to subtract the NLO($|q^2|<Q^2_0$) contribution from both the quark and gluon coefficient functions, $C^{\rm NLO}_q$ and $C^{\rm NLO}_g$. The formulae that come from this non-trivial calculation are given in the Appendix of Ref.~\cite{JMRTfinal}.  We use them to perform the numerical computations to obtain the NLO prediction after the subtractions. The result is shown in the lower plot in Fig.~\ref{fig:Capture}.  We now have perturbative stability; the NLO contribution becomes a much smaller correction to the LO prediction.

It should be emphasized that the asymptotics of the NLO amplitude is used only to determine the effective scale $\mu_F$. In all our further numerics we use the full expressions for the NLO amplitudes.

Throughout we have chosen the renormalization scale equal to the factorization scale, that is $\mu_R=\mu_F$.
The arguments are as follows. First, this corresponds to the BLM prescription~\cite{BLM};  such a choice eliminates from the NLO terms the contribution proportional to $\beta_0$ (that is, the term $\beta_0{\rm ln}(\mu_R^2/\mu_F^2)$ in eq. (3.95) of \cite{ISSK}. Second, following the discussion in~\cite{LHKR} for the analogous QED case, we note that the new quark loop insertion into the gluon propagator appears twice in the calculation. The part with scales $\mu <\mu_F$ is generated by the virtual component ($\propto \delta(1-z)$) of the LO splitting during DGLAP evolution, while the part with scales $\mu>\mu_R$ accounts for the running $\alpha_s$ behaviour obtained after the regularization of the ultraviolet divergence. In order not to miss some contribution and/or to avoid double counting we take the renormalization scale equal to the factorization scale, $\mu_R=\mu_F$.

\section{Conclusion}

We have shown that the bad perturbative convergence and the large sensitivity of the QCD prediction for exclusive $\J$ forward production can be avoided if (i) the factorization scale is chosen to be $\mu_F=m_c$ so that the double log  terms, $(\alpha_s{\rm ln}(1/\xi){\rm ln}(\mu_F^2))^n$, in the NLO coefficient functions are transferred to the LO contribution, and (ii) the $|q^2|<Q^2_0$ contribution is removed from the NLO coefficient functions to avoid double counting. These modifications provide reasonable accuracy for the NLO $\gamma p\to \J+p$ amplitude in the collinear $\MS$ factorization scheme, and open the possibility that data, for high precision exclusive production of $\J$ mesons in the forward direction, can be included in the global parton analyses to determine the low $x$ gluon PDF.

Exclusive $\Upsilon$ production can be predicted more reliably theoretically than $\J$ production, but there will be fewer experimental events.

\section{Acknowledgements}
We thank Ronan McNulty for valuable discussions. SPJ is supported by the Research Executive Agency (REA) of the European Union under the Grant Agreement PITN-GA2012316704 (HiggsTools).

\bibliographystyle{JHEP}
\providecommand{\href}[2]{#2}\begingroup\raggedright\endgroup

\end{document}